\documentclass[reprint,twocolumn,english,aps,prl]{revtex4-1}
\usepackage[T1]{fontenc}
\usepackage[latin9]{inputenc}
\usepackage{prettyref}
\usepackage{float}
\usepackage{amsmath}
\usepackage{graphicx}

\usepackage{babel}

\begin{document}

\maketitle

\textbf{Reply to Comment on ``Dislocation Structure and Mobility in hcp $^4$He''} \\
\\
In our recent Letter~\cite{LandinezBorda2016} we report path-integral Monte Carlo simulations considering the structure and mobility of basal-plane dislocations in hcp $^4$He and discuss the results in the context of (i) observations of giant plasticity and (ii) recent mass-flow junction experiments. In their Comment~\cite{KuklovComment2016} Kuklov and Svistunov argue that, (a) our discussion of part (ii) is misleading, (b) our results do not provide new insight into dislocation dissociation nor superfluidity of dislocation cores and (c) our calculations lack control of the numerical data. In the following paragraphs we discuss their assertions.

In (a) the authors of the Comment argue that our statement that ``the interpretation of recent mass flow experiments in terms of a network of 1D L{\"u}ttinger-liquid systems in the form of superfluid dislocation cores does not involve basal-plane dislocations'' is misleading. This assertion is likely to be due to a misunderstanding regarding the definition of basal-plane dislocations. In our Letter, we define basal-plane dislocations as those that have \emph{both} line direction \emph{and} their Burgers vector contained in the basal plane. This terminology is consistent with the fact that it is this set of dislocations that mediates basal slip, the dominant mechanism for plastic deformation in hcp $^4$He\cite{Paalanen1981,Haziot2013}. Our results imply that this class of dislocations, also responsible for the observed giant plasticity in hcp $^4$He, do not have superfluid cores and thus are not expected to be involved in the detected mass-flow phenomena. In this sense there is nothing misleading about the above-mentioned statement.

In (b), it is argued that Ref.~\cite{LandinezBorda2016} ``neither negates the results of'' their earlier work~\cite{Boninsegni2007,Soyler2009,Pollet2007,Pollet2008}, ``nor provides new insights into superfluidity of dislocations.'' In particular, they mention the lack of novelty in our reports of dislocation dissociation and the absence of superfluidity. We disagree with this point. With the exception of a referral to unpublished results in Ref.~\cite{Pollet2008} concerning the nonsuperfluid dissociation of the same basal-plane edge dislocation studied in our Letter, none of the dislocations discussed in Refs.~\cite{Boninsegni2007,Soyler2009,Pollet2007,Pollet2008} are of the basal-plane type according to our definition and thus are different from those we discuss.  

Item (c) questions technical issues of our simulations. In particular, it argues that the absence of superfluidity  is ``most likely'' a ``lack of control of the numerical data'' due to ergodicity issues in the used permutation-sampling algorithm. Indeed, this algorithm runs into trouble when the size of the computational cell becomes large. However, even though the numbers of atoms in the cells are large, their size \emph{along} the dislocation lines remain small, spanning $\sim 8$ atomic distances. In other words, the winding paths along the dislocation cores remain sufficiently short to enable sampling of such paths. Indeed, fleeting cycles of lengths of 6-10 were routinely detected and our observation of the absence of superfluidity is robust.

Finally, it is worth noting that our work also represents a novelty in the context of technical execution. It is well known that for atomistic simulations of dislocations to produce meaningful results, utmost care must be taken with the boundary conditions. In Ref.~\cite{LandinezBorda2016} this has been done rigorously, allowing the dislocation dipoles to fully relax in the isothermal-isostress ensemble and analyzing in detail the image effects due to the periodic boundary conditions. In the simulations described in Refs.~\cite{Boninsegni2007,Soyler2009}, on the other hand, this issue has not been treated on the same level. Specifically, single dislocations were modeled using cylinder-type computational cells featuring frozen atoms on the outside. It is an established fact~\cite{Sinclair1978,Rao1998} that such an approach introduces incompatibility stresses and that the failure to relieve them, for instance by using Green's function boundary conditions~\cite{Rao1998}, may result in incorrect dislocation core structures.\\

Edgar Josué Landinez Borda$^1$, Wei Cai$^2$ and Maurice de Koning$^3$ 

$^1$Lawrence Livermore National Laboratory,  7000 East Ave, Livermore, CA 94550.

$^2$Department of Mechanical Engineering, Stanford University, Stanford, CA 94305-4040.
 
$^3$Instituto de Física "Gleb Wataghin", Universidade Estadual de Campinas, UNICAMP, 13083-859, Campinas, São Paulo, Brazil. 

\bibliographystyle{apsrev4-1}
%

\end{document}